# Realizable Eddy Damped Markovian Anisotropic Closure for Turbulence and Rossby Wave Interactions

Jorgen S. Frederiksen [1]*  and Terence J. O'Kane[2]

1. CSIRO Environment, Aspendale, Melbourne, 3195, Australia; Jorgen.Frederiksen@csiro.au
2. CSIRO Environment, Hobart, 7004, Australia; Terence.O'Kane@csiro.au
* Correspondence: Jorgen.Frederiksen@csiro.au

**Abstract:** A realizable Eddy Damped Markovian Anisotropic Closure (EDMAC) is presented for the interaction of two-dimensional turbulence and transient waves such as Rossby waves. The structure of the EDMAC ensures that it is as computationally efficient as the Eddy Damped Quasi Normal Markovian (EDQNM) closure but unlike the EDQNM is guaranteed to be realizable in the presence of transient waves. Jack Herring's important contributions to laying the foundations of statistical dynamical closure theories of fluid turbulence are briefly reviewed. The topics covered include equilibrium statistical mechanics, Eulerian and quasi-Lagrangian statistical dynamical closure theories, and the statistical dynamics of interactions of turbulence with topography. The impact of Herring's work is described and placed in the context of related developments. Some of the further works that have built on Herring's foundations are discussed. The relationships between theoretical approaches employed in statistical classical and quantum field theories, and their overlap, are outlined. The seminal advances made by the pioneers in strong interaction fluid turbulence theory are put in perspective by comparing related developments in strong interaction quantum field theory.

## 1. Introduction

Jack Herring (1975) [1] developed a theory of two-dimensional (2D) anisotropic turbulence by generalizing the Eddy-Damped Quasi-Normal Markovian (EDQNM) closure that had been derived by Orszag (1970) [2] for three-dimensional (3D) homogeneous isotropic turbulence (HIT). The EDQNM was numerically implemented and studied for 2D HIT by Leith (1971) [3]. Herring's interest was in examining the relaxation of 2D homogeneous anisotropic turbulence (HAT) back to isotropy in comparison with the return of 3D HAT to HIT. The study was performed without the presence of transient Rossby waves. Indeed, incorporating transient waves in the EDQNM closure, while guaranteeing realizability, has been a long-standing problem that we will discuss in detail.

Our aim in this article is to present a variant of the EDQNM closure for HAT, which is realizable in the presence of transient waves. The study focuses on Rossby waves, although the same approach can be used for other waves. We call the model the realizable Eddy Damped Markovian Anisotropic Closure (EDMAC). The EDMAC model is as computationally efficient as the EDQNM and so expands the applicability Markovian closures to cater for transient waves at very little computational cost.

A second aim is to summarize some of Jack Herring's major achievements in laying the foundations of the statistical dynamical theory of turbulence. It is also to place his work in the context of related developments, to note the impacts of his work and how it has allowed further advances in this complex and difficult field.

The article is structured as follows. In Section 2, we review some of Jack Herring's major pioneering contributions to formulating the foundations of the statistical dynamical closure theory for fluid turbulence. His impact on the field is also discussed and his work placed in the context of related works and some of the further advances that have built on these foundations are discussed. A major aim of this article is also to present a generalization of the EDQNM closure that is guaranteed to be realizable for 2D HAT in the presence of transient waves like Rossby waves. This work thus extends the closures of Orszag [2] and Herring [1] and provides a resolution of a long-standing problem.

In Section 3, the dynamical equations for 2D HAT interacting with Rossby waves on a $\beta$ – plane are summarized. The corresponding direct numerical simulation equations are displayed in the spectral space of Fourier coefficients on the doubly periodic domain in Section 4. In Section 5, non-Markovian closures for 2D HAT are presented and three variants of corresponding Markovian closures with auxiliary evolution equations for the triad relaxation functions are derived in Section 6. The realizable EDMAC closure for 2D HAT interacting with transient Rossby waves is presented in Section 7. The EDMAC model is constructed to be as numerically efficient as the EDQNM closure since it has an analytical expression for the triad relaxation function like the EDQNM. Section 8 contains a few of Jorgen's personal reflections on Jack Herring and perspectives on strong interaction statistical field theories, and a comparison of progress in strong interaction fluid turbulence with that in hadron physics in quantum field theory. Our conclusions are summarized in Section 9. Appendix A establishes the conditions under which the real part of the EDMAC triad relaxation function is positive semi-definite and Appendix B presents the Langevin equation that underpins the EDMAC model and ensures that it is realizable for 2D HAT in the presence of transient Rossby waves.

## 2. Herring's Statistical Dynamical Theories and their Impacts and Extensions

Jack Herring made giant pioneering steps in laying the foundations of the statistical dynamical theories of fluid turbulence. In this Section, we briefly summarize some of his major achievements, the impacts of his works and some of the related and further developments that have occurred.

*2.1 Equilibrium Statistical Mechanics*

Equilibrium statistical mechanics is perhaps the simplest theoretical framework that gives some insight into the more complex phenomena of turbulence in the presence of forcing and dissipation. The appeal of equilibrium statistical mechanics is that the inviscid unforced equations of motion, with conservation laws like energy and enstrophy, have exact analytical solutions.

*2.1.1 Complete Statistical Mechanics Theories*

Herring (1977) [4] developed the statistical mechanics theory of two-dimensional flows over random topography with ensemble averages taken over both the flow fields and the topography. He formulated the canonical equilibrium solutions based on planar geometry spectral representations of the flow fields and topography. The aim was to provide guidance and understanding of the more complicated statistical dynamical closure theories that he formulated and solved numerically for forced dissipative turbulence over random topography. Herring's thermal equilibrium solutions for ensembles of random topography led on from the earlier statistical mechanics of point vortices by Onsager (1949) [5], Kraichnan's (1967) [6} and (1975) [7] planar geometry spectral solutions without topography, and the planar geometry statistical mechanics for quasigeostrophic flows over single realization topography by Salmon et al. (1976) [8].

These seminal works were extended to formulate canonical equilibrium theory in spherical geometry both with and without single realization topography [9,10]. Indeed, the relationships between equilibrium statistical mechanics solutions in planar geometry and on the sphere have been further elucidated and reviewed in this Special Issue dedicated to Jack Herring by Salmon and Pizzo [11].

Bretherton and Haidvogel (1976) [12] developed a different model for understanding turbulence over topography in which the dynamical solutions were minimum enstrophy states that are nonlinearly stable [13]. Frederiksen and Carnevale (1986) [14] established the relationships between canonical equilibrium states and minimum enstrophy states for barotropic flows over topography on the sphere and this was subsequently generalized to baroclinic flows [15,16]. The corresponding equivalence was formulated for barotropic flows over topography in planar geometry by Carnevale and Frederiksen (1987) [17]. They also considered the thermodynamic limit of infinite resolution and showed that then the canonical equilibrium state is statistically sharp and identical to the nonlinearly stable minimum enstrophy state. In their study it was also noted that for the continuum dynamics of fluids more general nonlinear stable states, than the minimum enstrophy states, are possible since an infinity of invariants exists in the inviscid case. Moreover, they pointed out that statistical mechanics theory can also be generalized to account for these invariants to be consistent with the many-invariant nonlinearly stable flows. The in principle and practical difficulty is of course developing a realizable numerical model with an infinity of invariants on which theoretical formulations and solutions can be based.

*2.1.2 Empirical Statistical Mechanics Theories*

Miller and collaborators [18,19] and Robert and collaborators [20,21] made attempts at formulating complete statistical mechanics theories with an infinity of invariants. However, shortcomings of these attempts have been noted in several studies. Chorin [22] and Turkington [23] pointed out that the lattice models of Miller et al. [19] always have a shortest scale while this is not the case for continuum fluids. Majda and Wang [24] regard the many-invariant approaches as empirical rather than complete statistical theories like the energy-enstrophy theories with underpinning realizable spectral models. More extensive discussions of statistical mechanics methods including recent works and applications are provided in the reviews in Refs. [24–27].

*2.2 Eulerian Statistical Dynamical Closure Theories*

The late 1950s to mid-1960s was a time of extraordinary advances in the theory of strong turbulence, one of the most difficult problems in classical physics. In the vanguard was the Eulerian Direct Interaction Approximation (DIA) closure theory of Kraichnan [28,29] for homogeneous isotropic turbulence (HIT). The DIA was based on formal, rather heuristic, renormalized perturbation theory and in the language of modern physics is a bare vertex approximation [30–32]. This was followed by the nonequilibrium steady state theory of Edwards [33] and the Self Consistent Field Theory (SCFT) of Herring [34,35] in steady state and time-dependent forms. Both the Edwards and Herring statistical closures for HIT are based on Liouville or Fokker-Planck formalisms and are original in their approaches. Carnevale and Frederiksen [36] and McComb [37] in this Special Issue in commemoration of Jack Herring compare these different formalisms at nonequilibrium steady states. Another Eulerian non-Markovian closure theory for HIT that was developed in the 1970s by McComb [38–40] is the Local Energy Transfer Theory (LET).

The three time-dependent Eulerian non-Markovian closure theories developed by Kraichnan, Herring and McComb are based on quite different physical reasoning and theoretical approaches. However, the final closure equations are in fact simply related. The theories are all bare vertex approximation theories. The theories all have the same single-time cumulant equation and differ just in how the two-time cumulant and response functions are treated. Kraichnan's DIA has separate equations for the response functions and two-time cumulants while both Herring's SCFT and McComb's LET effectively assume a Fluctuation Dissipation Theorem (FDT) between the response and two-time cumulant functions [41]. In principle the FDT should only be strictly valid in statistical mechanical equilibrium as described in Section 2.1. The SCFT and LET closures can be derived from the DIA by employing the *prior-time FDT* [37] (Equation (3.5)) defined by

$$C_{\mathbf{k}}(t,t') \equiv R_{\mathbf{k}}(t,t')C_{\mathbf{k}}(t',t') \tag{1}$$

for $t \geq t'$. Here, $C_{\mathbf{k}}(t,t')$ is the two-time spectral cumulant at wavenumber $\mathbf{k}$, $R_{\mathbf{k}}(t,t')$ is the response function and $C_{\mathbf{k}}(t',t')$ the prior time single-time cumulant. The SCFT and DIA have the same response function equation but the SCFT then calculates the two-time cumulant from Equation (1). On the other hand, the LET and the DIA have the same two-time cumulant equation with the LET determining the response function from Equation (1).

The DIA, SCFT and LET closures are quite skillful in capturing the large energy containing scales which is probably the most important property for geophysical fluids and plasmas. However, the focus of much of the statistical closure theory for HIT has been on the power law behaviour of the closures as well as their realizability. The realizability of the DIA closure was a major triumph but the power laws of the DIA, and the SCFT, were found to differ a little from the inertial range power laws of $k^{-\frac{5}{3}}$ for 3D turbulence and from the $k^{-3}$ enstrophy cascading inertial range for two-dimensional turbulence (Herring et al. 1974 [42]). Again, the power law fall-off of the LET closure and general performance are very similar to the DIA and SCFT closures for two-dimensional turbulence (Frederiksen and Davies 2000 [43]). The power law deficiencies of the DIA closure were ascribed by Kraichnan [29] to spurious sweeping effects of the small eddies by the large eddies.

The study of Herring et al. [42] established the properties of the DIA closure for two-dimensional HIT in comparison with direct numerical simulations (DNS) on the doubly periodic domain. The DNS used the discrete Fourier transform but for the sake of efficiency the DIA closure, as with most closure codes, was formulated for the continuum problem and used logarithmic discretization to reach high wavenumbers. Frederiksen and Davies [43] made a series of similar comparisons of the DIA, SCFT and LET closures for two-dimensional HIT with DNS and for a range of large-scale Reynolds numbers between $R_L \approx 50$ and $R_L \approx 4000$ that included experimental setups very similar to those of Herring et al. [32]. They used closures formulated on the same discrete spectral space as the DNS which involved a much larger computational task. It meant that all interactions were included in both closures and DNS and a direct comparison was possible. It was found that the discrete closures were in much better agreement with DNS at the low to moderate Reynolds numbers used by Herring et al. [42]. As well, the DIA, SCFT and LET closures had very similar performance and similar deficits in kinetic energy and palinstrophy at the smallest scales compared with DNS.

Reviews of the subsequent development of closures for HIT are given in Refs. [27,37–40,44–47].

*2.3 Quasi-Lagrangian Statistical Dynamical Closure Theories*

The recognition of the inertial range discrepancies of the Eulerian DIA compared with the observed power laws of 2D and 3D turbulence led to the development of quasi-Lagrangian closures for HIT by Kraichnan (1965) [48] and (1977) [49], Herring and Kraichnan (1979) [50], Kaneda (1981) [51] and Gotoh et al. (1988) [52]. This was a truly epic effort of theoretical and numerical model development. Unfortunately, the outcome was not wholly successful. Unlike the Eulerian DIA, which is independent of field variable formulation, or norm, the quasi-Lagrangian closures depend on whether the derivations use labelling time derivatives [48] or measuring time derivatives [51]. As well, the quasi-Lagrangian closures depend on the choice of field variable, or representative. This is effectively equivalent to specifying one two-state parameter, the labelling or measuring time, and one continuous parameter that determines the field variable or linear combinations of field variables.

The quasi-Lagrangian closures are still second order in perturbation theory and the aims have been to make transformations that avoid the spurious convection effects identified by Kraichnan [29]. They do not, however provide a fundamental solution to the vertex renormalization problem "which is the whole problem of strong turbulence" (Martin et al. 1973 [53]).

Kraichnan (1964) [54] had earlier recognized that the power law discrepancies of the Eulerian DIA could be overcome by cutting off the wavenumber interactions between the larger and smaller scales in the response functions and two-time cumulants. This results in a one parameter regularized DIA closure with empirical vertex renormalization depending on the cut-off parameter $\alpha$ (see Section 5.3). However, it turns out that $\alpha$ is only weakly dependent on whether the turbulence is 2D or 3D and on whether the turbulence is homogeneous or inhomogeneous. For 3D HIT the regularized DIA closure captures the DNS statistics closely, including the $k^{-\frac{5}{3}}$ power law, with $\alpha$ between 3 and 3.5 [54–55]. For 2D turbulence the regularized DIA (Frederiksen and Davies 2004 [56]) and regularized QDIA (O'Kane and Frederiksen 2004 [57]) have employed values of $\alpha$ between 4 and 6. Closures for 2D HIT have used $\alpha = 6$ to give good agreement with DNS including the $k^{-3}$ power law and for inhomogeneous closures $\alpha = 4$ has been used. However, the closures in the HIT simulations were used with repeated cumulant update restarts [56] after relatively short times to optimize computational resources, and that allowed a larger value of $\alpha$ to be used. We in fact expect that, with sufficiently long-time integrals in non-Markovian closures, taking $\alpha \approx 4$ may be a reasonable universal value for both 2D and 3D turbulence.

The quasi-Lagrangian closures do of course improve on the small-scale behaviour of the Eulerian closures but there are still significant differences from the statistics of DNS. These differences depend on both the formulations and on whether the turbulence is 2D or 3D. The differences also generally become more apparent at higher Reynolds numbers. For the case of 2D HIT, Frederiksen and Davies [56] compared their regularized DIA calculations with quasi-Lagrangian closure results based on both labelling time derivatives and measuring time derivatives and for several choices of field, or representative, formulations. Compared with the statistics of DNS the regularized DIA performed better than the abridged Lagrangian-history direct interaction (ALHDI) approximation, and the strain-based abridged Lagrangian-history direct interaction (SALHDI) approximation of Herring and Kraichnan [50], and than the Lagrangian renormalized approximation (LRA) of Gotoh et al. [52].

*2.4 Homogeneous Closures for Turbulent Flows over Topography*

Herring (1977) [58] generalized the Eulerian DIA and the Test Field Model (TFM) (Kraichnan 1971 [59]) to study the statistics of turbulent 2D flows over ensembles of random topography. He considered rotating flows on an $f$ – plane, so there was no differential rotation with latitude, or $\beta$ – effect, or propagating Rossby waves. The aim was to shed further light into the findings of Bretherton and Haidvogel [12] that slowly decaying turbulence over topography tends to progress through a sequence of minimum enstrophy states for a fixed energy (in the absence of the $\beta$ – effect).

Herring [58] was able to establish broad parameter ranges for which there was general agreement between his forced dissipative closure results and the exact static solutions for inviscid flows. The locking of the flow to the topography was determined for the DIA and TFM closures with very similar behaviour at the larger scales.

*2.5 Inhomogeneous Closures for Turbulent Flows over Topography*

Frederiksen (1999) [59] continued the work on understanding the effects of topography on 2D turbulence. He developed a generalization of the Eulerian DIA closure theory that applies to inhomogeneous turbulence over single realization topography. This quasi-diagonal direct interaction approximation (QDIA) has a statistical dynamical equation for the mean flow unlike homogeneous closure equations for which the mean flow is zero. The mean flow also couples to the single-time cumulant, the two-time cumulant and the response function that describe the turbulence. The QDIA expresses the off-diagonal elements of the cumulants and response functions, in Fourier space, in terms of the diagonal elements and the mean flow and topography. Consequently, the QDIA is only a few times more computationally demanding than the corresponding DIA for homogeneous turbulence.

The inhomogeneous QDIA closure theory has been generalized to multi-field classical systems including quasi-geostrophic and 3D turbulence [60] and to quantum field theories [32]. O'Kane and Frederiksen (2004) [61] implemented the closure for numerical studies in comparison with ensembles of DNS on an $f$–plane. Frederiksen and O'Kane (2005) [62] generalized the QDIA closure for the interaction of inhomogeneous turbulent flows with Rossby waves and topography on a $\beta$–plane and noted the remarkable agreement between the closure and large ensembles of DNS for low Reynolds number flows. The QDIA closure has been extensively applied in bare vertex form at lower Reynolds numbers and in regularized form with $\alpha = 4$ at higher Reynolds numbers. It has been used to study the dynamics of turbulence, Rossby wave and topography interactions, the predictability of blocking regime transitions, data assimilation, and extensively applied for developing subgrid scale parameterizations. Reviews of the literature are given in Refs. [44,63,64].

*2.6 Markovian Statistical Closure Theories without Waves*

The non-Markovian closures, the DIA, SCFT and LET for homogeneous turbulence, and the QDIA for inhomogeneous turbulence, present a large computational task that scales as $O(T^3)$ where $T$ is the length of the time integration. This can be improved to scaling like $O(T^2)$ by restarting the integrations periodically and using the updated three-point cumulant in the restart [43,61,65]. However, Markovian closures such as Orszag's (1970) [2] EDQNM closure are still very much more computationally efficient since they scale like $O(T)$. The EDQNM is a one parameter theory with the eddy damping specified analytically to satisfy inertial range power laws. It was first implemented numerically by Leith (1971) [3] for 2D HIT with a form equivalent to that given in our Section 7.1. Herring (1975) [1] generalized the EDQNM closure to develop a theory of 2D anisotropic turbulence without the presence of waves. The EDQNM has an underpinning Langevin equation representation, as noted by Leith [3] and Herring and Kraichnan [66], that guarantees realizability for HIT and HAT without transient waves.

The EDQNM in its most basic form is represented by just the single-time cumulant spectral equation. It can be arrived at from the DIA closure by invoking the *current-time FDT*

$$C_{\mathbf{k}}(t,t') \equiv R_{\mathbf{k}}(t,t') C_{\mathbf{k}}(t,t) \qquad (2)$$

for $t \geq t'$. As well an analytical form for the response function is assumed and the time-history integrals of the DIA are then performed analytically to determine a triad relaxation function that enters the cumulant equation (see Section 7.1).

In the presence of transient waves, it is possible for the EDQNM closure to have unphysical behaviour and blow up (Bowman et al. 1993 [67]). To avoid this the steady state form of the triad relaxation function has often then been used [46, 68–70] or modified quasi-Normal Markovian closures employed [71,72].

*2.7 Markovian Statistical Closure Theories with Waves*

Bowman et al. [67] made detailed analyses of the EDQNM and the problems with realizability that can occur in the presence of transient waves. In particular, in the standard formulation, when waves, such as drift-waves or Rossby waves, are included in the response function it is not possible to guarantee that the real part of the triad relaxation function will be non-negative. This is also the case if the prior time FDT in Equation (1) is used instead of the current time FDT in Equation (2). Bowman et al. [67] found that they could derive a Realizable Markovian Closure (RMC) by using a fluctuation dissipation theorem that involves both current and prior time cumulants and that we have termed the *correlation FDT*

$$C_{\mathbf{k}}(t,t') \equiv [C_{\mathbf{k}}(t,t)]^{\frac{1}{2}} R_{\mathbf{k}}(t,t') [C_{\mathbf{k}}(t',t')]^{\frac{1}{2}} \qquad (3)$$

for $t \geq t'$. Note though that while the EDQNM triad relaxation function has an analytical form for the RMC a time-dependent equation must be solved to determine it.

The works of Bowman et al. [67], Hu et al. [73] and Bowman and Krommes [74] describe the further development of realizable Markovian closures for homogeneous turbulence with transient waves, in which the triad relaxation functions are determined by auxiliary differential equations.

Markovian Inhomogeneous Closures (MICs) were developed and tested against large ensembles of DNS by Frederiksen and O'Kane [44,75]. They started their formulations with the inhomogeneous QDIA closure for turbulent 2D flow interacting with Rossby waves and topography and employed the three versions of the FDT that they combined as:

$$C_{\mathbf{k}}(t,t') \equiv [C_{\mathbf{k}}(t,t)]^{1-X} R_{\mathbf{k}}(t,t') [C_{\mathbf{k}}(t',t')]^{X} \qquad (4)$$

for $t \geq t'$ and $C_{\mathbf{k}}(t,t') = C_{-\mathbf{k}}(t',t)$ for $t' > t$. Here, $X = 0$ for the current-time FDT used for the EDQNM of Orszag {1970}, $X = \frac{1}{2}$ for the correlation FDT used for the RMC of Bowman et al. [67], $X = 1$ for the prior-time FDT used for the SCFT of Herring [35] and the LET of McComb [38,39].

All the MICs developed by Frederiksen and O'Kane [44,75] performed remarkably well compared with the DNS ensembles in low Reynolds number numerical experiments. However, it is of course desirable to be sure that the Markovian closures employed will be realizable under all circumstances as is the case for the formulations using the correlation FDT ($X = \frac{1}{2}$). Determining the relaxation functions through time integration of differential equations is nevertheless a considerable computational overhead. Frederiksen and O'Kane [75] therefore developed the Eddy Damped Markovian Inhomogeneous Closure (EDMIC) that generalizes the EDQNM to inhomogeneous flows. The EDMIC has analytical forms for the relaxation functions and is realizable under the same conditions as the EDQNM.

A major aim of this article is to formulate a realizable Eddy Damped Markovian Anisotropic Closure (EDMAC) with analytical triad relaxation function in the presence of transient Rossby waves that thus generalizes the EDQNM.

*2.8 Classical and Quantum Statistical Field Theory Formalisms*

Herring's [34,35] theoretical approach to deriving the statistical dynamical closure equations based on Liouville of Fokker-Planck formalisms was original. It was different from the approaches that had become standard in quantum field theory since the remarkable success of renormalized perturbation theory in quantum electrodynamics by Feynman, Schwinger and Tomonaga in the mid-20[th] century [32]. This is also the case for the works of Edwards [33], McComb [38,39], and the published work of Kraichnan [28,29]. However, Martin et al. [53] (Footnote 11) note "It seems that Kraichnan's rules for calculating the renormalized vertices to a given order generate the quantities which are given exact nonperturbative definitions here. We are grateful to Dr. Kraichnan for providing us with old unpublished notes …".

Wyld [30] and Lee [31] reconstructed Kraichnan's Eulerian DIA closure through renormalized perturbation theory with a diagrammatic representation similar to Feynman diagrams. Lee considered magneto-hydrodynamics to sixth order and pointed out that Wyld had mistakenly replaced the bare propagator by the renormalized propagator in some terms in his fourth order representation. The DIA in these works appears as a bare vertex approximation in second order renormalized perturbation theory.

Martin et al. [53] (hereafter MSR) generalized the Schwinger-Dyson functional operator approach to classical statistical dynamics by introducing an adjoint operator that generates the response function. They also introduced an associated non-Hermitian Hamiltonian, in terms of the field variable and the adjoint operator, and a generating functional from which the mean field, two-point cumulants, response functions, self-energies and vertex functions can be derive through functional differentiation.

The MSR formalism was expanded by Rose [76] to include random forcing including non-Gaussian noise and non-Gaussian initial conditions. The full power of renormalized perturbation theory was achieved by the reformulation of the MSR approach through the Feynman path integral formalism [77] by Phythian [78] and most generally by Jensen [79].

Berera et al. [80] made an important reconciliation between the Wyld diagrammatic approach at fourth order and the MSR formalism. As further discussed by McComb [37] they corrected some errors in both the Wyld and MSR formalisms to show the consistency between the diagrammatic approach and the functional method, as would be expected.

The Schwinger-Dyson functional formalism was designed for developing a statistical theory of scattering from asymptotic 'in' states to 'out' states. This is analogous to the steady state theories of Edwards [33] and Herring [34] with the fluctuation dissipation theorem (FDT) imposed. The time-dependent nonequilibrium statistical dynamical theories are more complex, do not satisfy the FDT <u>exactly</u>, and that is why MSR needed to introduce the adjoint operator to generate the separate response function. The adjoint operator incidentally arises naturally in the path integral generalization [78,79]. In fact, the MSR (1973) [53] formalism and its path integral generalizations for classical systems are much more similar to the closed time path (CTP) formalism of Schwinger (1961) [81] and Keldysh (1965) [82] for time-dependent nonequilibrium quantum field theories. The earlier Schwinger-Keldysh CTP theory contains all the aspects of adjoint operators, resulting in operator doubling, Pauli matrices, two-point matrix Greens functions consisting of cumulants and response functions and matrix self-energies that appear in the MSR formalism. Indeed, it is surprising that the Schwinger-Keldysh formalism was not referenced by MSR.

Analyses of the relationships between the Schwinger-Keldysh CTP quantum field theory formalism and the MSR classical field theory approach are described in the works of Cooper et al. [83] and Blagoev et al. [84]. Frederiksen [32] presents a parallel development of classical and quantum statistical field theories based on these formalisms, and the path integral generalization, to include non-Gaussian noise, non-Gaussian initial conditions, and quantum effects. It is shown that the approaches are equivalent and that the classical approach can be applied to generate the second order statistical equations for quantum systems provided a suitable non-Gaussian noise term is included. For the Klein-Gordon equation with interaction Lagrangian of $g\phi^3$ the quantum effects correspond to a pure skewness non-Gaussian noise and for $\lambda\phi^4$ it corresponds to pure kurtosis. Thus, the differences between classical and quantum field theories manifest themselves through an additional quantum self-energy term proportional to Planck's constant squared, and in the initial conditions.

Frederiksen [32] also developed the inhomogeneous QDIA closure for quantum field theories. For both classical and quantum field theories the QDIA closure is computationally tractable. This is because the off-diagonal elements of the two-point cumulants and response functions – the propagators – in Fourier space are represented in terms of the diagonal cumulants and response functions and the mean fields (and possibly topography).

### 3. Two-dimensional Barotropic Flows on a $\beta$–plane

In this study we focus on 2D turbulent flows in planar geometry and with differential rotation in the latitudinal direction that generates the $\beta$–effect. The equations of motion are most conveniently described by a single equation for the vorticity, $\zeta$. The vorticity is of course the Laplacian of the streamfunction $\psi$ whose gradient in turn is related to the zonal and meridional wind fields. Throughout this article we develop our results on the doubly periodic domain $0 \leq x \leq 2\pi, 0 \leq y \leq 2\pi$ with $\mathbf{x} = (x, y)$.

The barotropic vorticity equation is given by

$$\frac{\partial \zeta}{\partial t} = -J(\psi, \zeta + \beta y) + \hat{v}_0 \nabla^2 \zeta + f^0. \tag{5}$$

Here, the Jacobian is

$$J(\psi, \zeta) = \frac{\partial \psi}{\partial x}\frac{\partial \zeta}{\partial y} - \frac{\partial \psi}{\partial y}\frac{\partial \zeta}{\partial x} \tag{6}$$

and the vorticity $\zeta$ is the Laplacian of the streamfunction

$$\zeta = \nabla^2 \psi \equiv \left(\frac{\partial^2}{\partial x^2} + \frac{\partial^2}{\partial y^2}\right)\psi. \tag{7}$$

In these equations, $\beta$ denotes the beta-effect, $f^0$ specifies any external forcing and $\hat{v}_0$ is the viscosity. The dissipation in Equation (5) is represented by the Laplacian but we shall in fact also consider higher order dissipation operators so that in spectral space $\hat{v}_0$ depends on the wavenumber.

Rossby waves, and superpositions of Rossby waves, proportional to $\exp i(\mathbf{k}.\mathbf{x} - \omega_\mathbf{k} t)$ are solutions to Equation (5) (in the absence of forcing, topography, and viscosity). The Rossby wave frequency satisfies the dispersion relationship

$$\omega_\mathbf{k} = \omega_\mathbf{k}^\beta = -\frac{\beta k_x}{k^2}. \tag{8}$$

### 4. Dynamical Equations in Fourier Space

Our analysis and theoretical developments will be performed in Fourier space with each of the fields having a spectral representation similar to that of the vorticity:

$$\zeta(\mathbf{x}, t) = \sum_{\mathbf{k} \in \mathbf{R}} \zeta_\mathbf{k}(t) \exp(i \mathbf{k}.\mathbf{x}), \tag{9}$$

where

$$\zeta_\mathbf{k}(t) = \frac{1}{(2\pi)^2} \int_o^{2\pi} d^2\mathbf{x} \zeta(\mathbf{x}, t) \exp(-i \mathbf{k}.\mathbf{x}). \tag{10}$$

Here, $\mathbf{x} = (x, y)$, $\mathbf{k} = (k_x, k_y)$, $k = (k_x^2 + k_y^2)^{1/2}$. The reality of the physical space fields implies that in spectral space $\zeta_{-\mathbf{k}} = \zeta_\mathbf{k}^*$. The summations in Equation (9) are over the domain $\mathbf{R}$ which is a circular wavenumber domain excluding the origin $\mathbf{0}$. In spectral space the resulting dynamical equation is:

$$\left(\frac{\partial}{\partial t} + v_0(\mathbf{k})k^2\right)\zeta_\mathbf{k}(t) = \sum_{\mathbf{p} \in \mathbf{R}}\sum_{\mathbf{q} \in \mathbf{R}} \delta(\mathbf{k}, \mathbf{p}, \mathbf{q}) K(\mathbf{k}, \mathbf{p}, \mathbf{q}) \zeta_{-\mathbf{p}}(t)\zeta_{-\mathbf{q}}(t) + f_\mathbf{k}^0(t). \tag{11}$$

The generalized delta function is defined by $\delta(\mathbf{k}, \mathbf{p}, \mathbf{q}) = 1$ if $\mathbf{k} + \mathbf{p} + \mathbf{q} = 0$ and 0 if $\mathbf{k} + \mathbf{p} + \mathbf{q} \neq 0$. It is also convenient to define the complex $v_0(\mathbf{k})k^2$ that represents both the viscosity and the Rossby wave frequency $\omega_\mathbf{k}$:

$$\nu_0(\mathbf{k})k^2 = \hat{\nu}_0(k)k^2 + i\omega_{\mathbf{k}}. \tag{12}$$

The Rossby wave frequency is defined in Equation (8). We have also generalized the form of the viscosity $\hat{\nu}_0 \to \hat{\nu}_0(k)$ to allow for more general dissipation operators in Equation (5). The interaction coefficient $K(\mathbf{k},\mathbf{p},\mathbf{q})$ is given by

$$K(\mathbf{k},\mathbf{p},\mathbf{q}) = \tfrac{1}{2}[p_x q_y - p_y q_x](p^2 - q^2)/p^2 q^2. \tag{13}$$

## 5. Eulerian Non-Markovian Statistical Dynamical Closures

Statistical dynamical closures encapsulate the dynamics of infinite ensembles of characteristic flows. Our focus here is on 2D flows described by the DNS in Equation (5) (in physical space) and Equation (11) (in Fourier space). For a given member making up the ensemble, the particular flow field, here given by the vorticity spectral coefficient $\zeta_{\mathbf{k}}(t)$ at wavenumber $\mathbf{k}$, is conveniently represented by its mean $<\zeta_{\mathbf{k}}(t)> \equiv \overline{\zeta}_{\mathbf{k}}(t)$ and deviation from the ensemble mean $\tilde{\zeta}_{\mathbf{k}}(t)$. For homogeneous turbulence, with zero mean flow, we have

$$<\zeta_{\mathbf{k}}(t)> \equiv \overline{\zeta}_{\mathbf{k}}(t) = 0 \ ; \ \zeta_{\mathbf{k}}(t) = \tilde{\zeta}_{\mathbf{k}}(t). \tag{14}$$

The spectral equation for $\tilde{\zeta}_{\mathbf{k}}$ is thus given by Equation (11) with $\zeta_{\mathbf{k}}(t) \to \tilde{\zeta}_{\mathbf{k}}(t)$. Given that the mean field is zero, this also means that the mean forcing is zero and that

$$<f_{\mathbf{k}}^0(t)> \equiv \overline{f}_{\mathbf{k}}^0(t) = 0 \ ; f_{\mathbf{k}}^0(t) = \tilde{f}_{\mathbf{k}}^0(t). \tag{15}$$

*5.1 The DIA Closure for Homogeneous Turbulence*

A simple formal derivation of the DIA closure equations for homogeneous 2D turbulence is given by Frederiksen [85]. The DIA closure consists of coupled equations for the two-time two-point cumulant

$$C_{\mathbf{k}}(t,t') = <\tilde{\zeta}_{\mathbf{k}}(t)\tilde{\zeta}_{-\mathbf{k}}(t')> \tag{16}$$

and the ensemble average response function

$$R_{\mathbf{k}}(t,t') = \left\langle \tilde{R}_{\mathbf{k}}(t,t') \right\rangle. \tag{17}$$

Here, the response function for an individual disturbance is

$$\tilde{R}_{\mathbf{k}}(t,t') = \frac{\delta \tilde{\zeta}_{\mathbf{k}}(t)}{\delta \tilde{f}_{\mathbf{k}}^0(t')}, \tag{18}$$

where $\delta$ denotes the functional derivative. The response function measures the change in the individual field $\tilde{\zeta}_{\mathbf{k}}(t)$ at time $t$ due to an infinitesimal change in the forcing $\tilde{f}_{\mathbf{k}}^0(t')$ at the earlier time $t'$.

We can obtain the equation for the two-time cumulant $C_{\mathbf{k}}$ from Equation (11), by multiplying each term by $\tilde{\zeta}_{-\mathbf{k}}(t')$ and averaging. This gives the cumulant equation

$$\left(\frac{\partial}{\partial t} + \nu_0(\mathbf{k})k^2\right)C_{\mathbf{k}}(t,t')$$
$$= \sum_{\mathbf{p}}\sum_{\mathbf{q}} \delta(\mathbf{k},\mathbf{p},\mathbf{q})K(\mathbf{k},\mathbf{p},\mathbf{q})<\tilde{\zeta}_{-\mathbf{p}}(t)\,\tilde{\zeta}_{-\mathbf{q}}(t)\,\tilde{\zeta}_{-\mathbf{k}}(t')> + <\tilde{f}_{\mathbf{k}}^0(t)\tilde{\zeta}_{-\mathbf{k}}(t')> \tag{19}$$

where $t > t'$ and $C_{\mathbf{k}}(t,t') = C_{-\mathbf{k}}(t',t)$ for $t' > t$. It can also be shown that

$$<\tilde{f}_{\mathbf{k}}^0(t)\tilde{\zeta}_{-\mathbf{k}}(t')> = \int_{t_0}^{t'} ds\, F_{\mathbf{k}}^0(t,s)R_{-\mathbf{k}}(t',s) \tag{20}$$

where $t_0$ is the initial time and

$$F_{\mathbf{k}}^0(t,s) = <\tilde{f}_{\mathbf{k}}^0(t)\tilde{f}_{\mathbf{k}}^{0*}(s)>. \tag{21}$$

Thus, the two-point cumulant equation is coupled to the response function through the random forcing term but, also it turns out, through the closure for the three-point cumulant in Equation (19). When this closure is performed [85] the two-time cumulant equation becomes

$$\left(\frac{\partial}{\partial t}+\nu_0(\mathbf{k})k^2\right)C_\mathbf{k}(t,t')+\int_{t_0}^{t}ds\,\eta_\mathbf{k}(t,s)C_{-\mathbf{k}}(t',s)$$
$$=\int_{t_0}^{t'}ds\,(S_\mathbf{k}(t,s)+F_\mathbf{k}^0(t,s))R_{-\mathbf{k}}(t',s) \quad (22)$$

where $t > t'$ with $C_\mathbf{k}(t,t') = C_{-\mathbf{k}}(t',t)$ for $t' > t$. In a similar way the response function equation can be derived [85] as

$$\left(\frac{\partial}{\partial t}+\nu_0(\mathbf{k})k^2\right)R_\mathbf{k}(t,t')+\int_{t'}^{t}ds\,\eta_\mathbf{k}(t,s)R_\mathbf{k}(s,t') = \delta(t-t') \quad (23)$$

for $t \geq t'$ and $R_\mathbf{k}(t,t') = 1$.

In Equations (22) and (23)

$$\eta_\mathbf{k}(t,s) = -4\sum_\mathbf{p}\sum_\mathbf{q}\delta(\mathbf{k},\mathbf{p},\mathbf{q})K(\mathbf{k},\mathbf{p},\mathbf{q})K(-\mathbf{p},-\mathbf{q},-\mathbf{k})R_{-\mathbf{p}}(t,s)C_{-\mathbf{q}}(t,s), \quad (24)$$

and

$$S_\mathbf{k}(t,s) = 2\sum_\mathbf{p}\sum_\mathbf{q}\delta(\mathbf{k},\mathbf{p},\mathbf{q})K(\mathbf{k},\mathbf{p},\mathbf{q})K(-\mathbf{k},-\mathbf{p},-\mathbf{q})C_{-\mathbf{p}}(t,s)C_{-\mathbf{q}}(t,s), \quad (25)$$

The above two terms are known in field theory as self-energies. They modify or renormalize the damping or forcing in the two-point cumulant and response function equations. The term $\eta_\mathbf{k}(t,s)$ is the nonlinear damping that appears in the two-point cumulant and response function equations. The nonlinear noise term, $S_\mathbf{k}(t,s)$, renormalizes the bare noise spectrum $F_\mathbf{k}^0(t,s)$ in the two-point cumulant equation. The noise terms are positive semi-definite.

The system of statistical dynamical equations is finally closed by the equation for the single-time two-point cumulant:

$$\left(\frac{\partial}{\partial t}+2\,\text{Re}\,\nu_0(\mathbf{k})k^2\right)C_\mathbf{k}(t,t)+2\,\text{Re}\int_{t_0}^{t}ds\,\eta_\mathbf{k}(t,s)C_{-\mathbf{k}}(t,s)$$
$$=2\,\text{Re}\int_{t_0}^{t}ds\,(S_\mathbf{k}(t,s)+F_\mathbf{k}^0(t,s))R_{-\mathbf{k}}(t,s) \quad (26)$$

where the initial conditions $C_\mathbf{k}(t_0,t_0)$ are to be specified.

*5.2 The SCFT and Let Closures for Homogeneous Turbulence*

McComb [37] has discussed in detail the historical developments of the SCFT closure of Herring and the LET closure of McComb. However, as noted in the Introduction, and in Ref. [43] the SCFT and LET closures can be obtained from the DIA by imposing the prior-time FDT. The three closures have in common the single-time cumulant prognostic in Equation (26). The SCFT also has the same expression as the DIA for the evolution of the response function in Equation (23) while the LET and DIA two-time cumulants are both determined by Equation (22). For the SCFT the two-time cumulant is then determined by the prior-time FDT in Equation (1) and for the LET the response function is instead obtained by this FDT.

*5.3 Regularized non-Markovian Closures for Homogeneous Turbulence*

Kraichnan [29] attributed the incorrect inertial ranges of the Eulerian DIA to spurious convection (advection) effects of the large eddies on the small-scale eddies. He showed that this could be overcome by restricting the ranges of interactions in the response function and two-time cumulant equations. He argued that this approach was analogous to using a quasi-Lagrangian formulation. We see the procedure as a regularization that corresponds to an empirical vertex renormalization. Specifically, we define

$$\breve{K}(\mathbf{k},\mathbf{p},\mathbf{q}) = \theta(p - k/\alpha)\theta(q - k/\alpha)K(\mathbf{k},\mathbf{p},\mathbf{q}), \tag{27}$$

where $\alpha$ is a wavenumber cut-off parameter and $\theta$ is the Heaviside step function. As noted in the Introduction, $\alpha$ only depends weakly on whether the turbulence is 2D or 3D or whether it is homogeneous or inhomogeneous. The regularized non-Markovian closures are obtained by replacing the interaction coefficient $K(\mathbf{k},\mathbf{p},\mathbf{q})$ by $\breve{K}(\mathbf{k},\mathbf{p},\mathbf{q})$ in Equation (23) for the response function and in Equation (22) for the two-time cumulant, but not in Equation (26) for the single-time cumulant.

## 6. Statistical Dynamical Equations for Markovian Anisotropic Closures

Next, we develop the theory of Markovian Anisotropic Closures (MACs) that consist of the single-time cumulant equation and an auxiliary differential equation for the evolution of the triad relaxation function. The MACs are designed to describe the statistical dynamics of anisotropic turbulence and one of the variants is guaranteed realizability even in the presence of transient Rossby waves. Based on the non-Markovian DIA in Section 5 we formulate three variants of the MACs that we refer to as $MAC^X$. Here, the superscript $X$ relates to that used in the combined FDT in Equation (4) with $X = 0$ being the current-time FDT, $X = \frac{1}{2}$ correlation FDT, and $X = 1$ the prior-time FDT.

The single-time two-point cumulant equation for the DIA, SCFT and LET closures is the same and can be written in the following form:

$$\frac{\partial}{\partial t}C_{\mathbf{k}}(t,t) + 2\,\mathrm{Re}[N_{\mathbf{k}}^{\eta}(t) + N_{\mathbf{k}}^{0}] = 2\,\mathrm{Re}[F_{\mathbf{k}}^{S}(t) + F_{\mathbf{k}}^{0}(t)] \tag{28}$$

where $C_{\mathbf{k}}(t,t)$ is real. The $F_{\mathbf{k}}(t)$ and $N_{\mathbf{k}}(t)$ functions have the following expressions:

$$F_{\mathbf{k}}^{S}(t) = 2\sum_{\mathbf{p}}\sum_{\mathbf{q}}\delta(\mathbf{k},\mathbf{p},\mathbf{q})K(\mathbf{k},\mathbf{p},\mathbf{q})K(-\mathbf{k},-\mathbf{p},-\mathbf{q})\Delta(-\mathbf{k},-\mathbf{p},-\mathbf{q})(t), \tag{29}$$

with

$$\Delta(-\mathbf{k},-\mathbf{p},-\mathbf{q})(t) = \int_{t_0}^{t}ds R_{-\mathbf{k}}(t,s)C_{-\mathbf{p}}(t,s)C_{-\mathbf{q}}(t,s), \tag{30}$$

and

$$F_{\mathbf{k}}^{0}(t) = \int_{t_0}^{t}ds F_{\mathbf{k}}^{0}(t,s)R_{-\mathbf{k}}(t,s). \tag{31}$$

Also

$$N_{\mathbf{k}}^{\eta}(t) = -4\sum_{\mathbf{p}}\sum_{\mathbf{q}}\delta(\mathbf{k},\mathbf{p},\mathbf{q})K(\mathbf{k},\mathbf{p},\mathbf{q})K(-\mathbf{p},-\mathbf{q},-\mathbf{k})\Delta(-\mathbf{p},-\mathbf{q},-\mathbf{k})(t), \tag{32}$$

and

$$N_{\mathbf{k}}^{0}(t) = \nu_{0}(\mathbf{k})k^{2}C_{\mathbf{k}}(t,t) = D_{\mathbf{k}}^{0}C_{\mathbf{k}}(t,t), \tag{33}$$

where

$$D_{\mathbf{k}}^{0} = \nu_{0}(\mathbf{k})k^{2}. \tag{34}$$

We now apply the FDTs in Equation (4) to simplify the nonlinear noise and damping terms in Equations (29) and (32). The time history integrals can then be expressed by relaxation functions $\Theta^X$. The expressions for the relaxation functions can in turn be expressed through time dependent differential equations. This then effects the Markovianization with the single-time cumulant equation augmented by the differential equations for $\Theta^X$. Thus,

$$F_{\mathbf{k}}^{S}(t) = 2\sum_{\mathbf{p}}\sum_{\mathbf{q}} \delta(\mathbf{k},\mathbf{p},\mathbf{q})K(\mathbf{k},\mathbf{p},\mathbf{q})K(-\mathbf{k},-\mathbf{p},-\mathbf{q}) \\ \times C_{-\mathbf{p}}^{1-X}(t,t)C_{-\mathbf{q}}^{1-X}(t,t)\Theta^X(-\mathbf{k},-\mathbf{p},-\mathbf{q})(t), \quad (35)$$

with

$$\Theta^X(-\mathbf{k},-\mathbf{p},-\mathbf{q})(t) = \int_{t_0}^{t} ds R_{-\mathbf{k}}(t,s)R_{-\mathbf{p}}(t,s)R_{-\mathbf{q}}(t,s)C_{-\mathbf{p}}^{X}(s,s)C_{-\mathbf{q}}^{X}(s,s). \quad (36)$$

Also

$$N_{\mathbf{k}}^{\eta}(t) = D_{\mathbf{k}}^{\eta}(t)C_{\mathbf{k}}(t,t), \quad (37)$$

with

$$D_{\mathbf{k}}^{\eta}(t) = -4\sum_{\mathbf{p}}\sum_{\mathbf{q}} \delta(\mathbf{k},\mathbf{p},\mathbf{q})K(\mathbf{k},\mathbf{p},\mathbf{q})K(-\mathbf{p},-\mathbf{q},-\mathbf{k}) \\ \times C_{-\mathbf{q}}^{1-X}(t,t)C_{-\mathbf{k}}^{-X}(t,t)\Theta^X(-\mathbf{p},-\mathbf{q},-\mathbf{k})(t). \quad (38)$$

The single-time cumulant equation then simplifies to

$$\left(\frac{\partial}{\partial t} + 2\operatorname{Re}(D_{\mathbf{k}}^{r}(t))\right)C_{\mathbf{k}}(t,t) = 2\operatorname{Re}(F_{\mathbf{k}}^{r}(t)) \quad (39)$$

and the response function equation becomes

$$\frac{\partial}{\partial t} R_{\mathbf{k}}(t,t') + D_{\mathbf{k}}^{r}(t)R_{\mathbf{k}}(t,t') = \delta(t-t'). \quad (40)$$

Here,

$$D_{\mathbf{k}}^{r}(t) = D_{\mathbf{k}}^{0} + D_{\mathbf{k}}^{\eta}(t) \; ; \; F_{\mathbf{k}}^{r}(t) = F_{\mathbf{k}}^{0}(t) + F_{\mathbf{k}}^{S}(t) \quad (41)$$

define the renormalized dissipation operator $D_{\mathbf{k}}^{r}(t)$ and the renormalized stochastic force $F_{\mathbf{k}}^{r}(t)$.

The integral form for the relaxation functions $\Theta^X$ in Equation (36) can be replaced by differential equations since the response functions are simplified in Equation (40). The ordinary differential equation for $\Theta^X$ is:

$$\frac{\partial}{\partial t}\Theta^X(\mathbf{k},\mathbf{p},\mathbf{q})(t) + (D_{\mathbf{k}}^{r}(t) + D_{\mathbf{p}}^{r}(t) + D_{\mathbf{q}}^{r}(t))\Theta^X(\mathbf{k},\mathbf{p},\mathbf{q})(t) \\ = C_{\mathbf{p}}^{X}(t,t)C_{\mathbf{q}}^{X}(t,t) \quad (42)$$

with $\Theta^X(\mathbf{k},\mathbf{p},\mathbf{q})(0) = 0$ and $D_{\mathbf{k}}^{r}$ given in Equation (41).

The three MACs with $X = 0, \frac{1}{2}, 1$ are specified by Equation (39) for the single-time cumulant $C_{\mathbf{k}}(t,t)$, together with Equation (42) for the relaxation function $\Theta^X(\mathbf{k},\mathbf{p},\mathbf{q})(t)$. The MACs with $X = 0, \frac{1}{2}, 1$ are all realizable in the absence of transient waves like Rossby waves and the variant with $X = \frac{1}{2}$ can also be shown to be realizable even in the presence of such waves [67].

We show in the next Section that it is possible to derive a Markovian closure with analytical representation of the triad relaxation function, which is therefore even more computationally efficient, from the MAC with $X = 0$. Note that when $X = 0$ Equation (39) for the single-time cumulant becomes:

$$\left(\frac{\partial}{\partial t} + 2\hat{v}_0(k)k^2\right)C_{\mathbf{k}}(t,t)$$
$$= 8\sum_{\mathbf{p}}\sum_{\mathbf{q}}\delta(\mathbf{k},\mathbf{p},\mathbf{q})K(\mathbf{k},\mathbf{p},\mathbf{q})K(\mathbf{p},\mathbf{q},\mathbf{k})\operatorname{Re}\Theta^0(\mathbf{k},\mathbf{p},\mathbf{q})(t) \qquad (43)$$
$$\times C_{\mathbf{q}}(t,t)\bigl[C_{\mathbf{k}}(t,t) - C_{\mathbf{p}}(t,t)\bigr].$$

Here, we have used the properties of the interaction coefficients that $K(\mathbf{k},\mathbf{p},\mathbf{q}) = K(\mathbf{k},\mathbf{q},\mathbf{p})$, and $K(\mathbf{k},\mathbf{p},\mathbf{q}) + K(\mathbf{p},\mathbf{q},\mathbf{k}) + K(\mathbf{q},\mathbf{k},\mathbf{p}) = 0$, the fact that the single-time cumulants are real, and the symmetry properties of the triad relaxation functions $\Theta^0$.

## 7. Realizable Eddy-Damped Markovian Anisotropic Closure

The MAC closures with $X = 0$, in Equations (36) and (43), can be simplified in a similar way to the derivation of the EDQNM closure [3,67–69]. In this Section, we establish the EDMAC model which is a suitable realizable generalization of the EDQNM for homogeneous turbulent flows interacting with transient Rossby waves. We seek to replace the differential equation in Equation (42) with an analytical parameterized expression that generalizes that used for the EDQNM model.

Firstly we note that the solution to the response function differential equation in Equation (40) is

$$R_{\mathbf{k}}(t,t') = \exp\left(-\int_{t'}^{t} ds\, D_{\mathbf{k}}^r(s)\right) \qquad (44)$$

where $D_{\mathbf{k}}^r$ is defined in Equation (41). Thus, from Equation (36) the triad relaxation function with $X = 0$ simplifies to

$$\Theta^0(\mathbf{k},\mathbf{p},\mathbf{q})(t) = \int_{t_0}^{t} dt' \exp\left(-\int_{t'}^{t} ds\left[D_{\mathbf{k}}^r(s) + D_{\mathbf{p}}^r(s) + D_{\mathbf{q}}^r(s)\right]\right). \qquad (45)$$

Next, we make the Markov approximation for the $D_{\mathbf{k}}^r$ such that $D_{\mathbf{k}}^r(s) \to D_{\mathbf{k}}^r(t)$. The generalized dissipation terms can therefore be taken outside the integrals in Equation (44). Thus,

$$R_{\mathbf{k}}(t,t') = \exp\left(-D_{\mathbf{k}}^r(t)(t-t')\right), \qquad (46)$$

and

$$\Theta^0(\mathbf{k},\mathbf{p},\mathbf{q})(t) = \int_{t_0}^{t} dt' \exp\left(-\left[D_{\mathbf{k}}^r(t) + D_{\mathbf{p}}^r(t) + D_{\mathbf{q}}^r(t)\right](t-t')\right)$$
$$= \frac{1 - \exp\left(-\left[D_{\mathbf{k}}^r(t) + D_{\mathbf{p}}^r(t) + D_{\mathbf{q}}^r(t)\right](t-t_0)\right)}{\left[D_{\mathbf{k}}^r(t) + D_{\mathbf{p}}^r(t) + D_{\mathbf{q}}^r(t)\right]}. \qquad (47)$$

have simpler analytical forms.

*7.1 Analytical Triad Relaxation Function for EDQNM*

In the EDQNM the only prognostic equation is Equation (43) for the second order cumulant $C_{\mathbf{k}}(t,t)$. Moreover, the eddy damping $D_{\mathbf{k}}^\eta$ that appears through $D_{\mathbf{k}}^r = D_{\mathbf{k}}^\eta + D_{\mathbf{k}}^0$ in the triad relaxation function in Equation (47) is generally specified by an analytical form that is consistent with the $k^{-3}$ enstrophy cascading inertial range:

$$D_{\mathbf{k}}^\eta(t) \to \mu_{\mathbf{k}}^{eddy}(t) = \gamma\left[k^2 C_{\mathbf{k}}(t,t)\right]^{\frac{1}{2}}. \qquad (48)$$

Integral forms over wavenumbers have also been used for the eddy damping [86] and our arguments here and in Section 7.2 apply equally for those forms. We note that $C_\mathbf{k}(t,t)$ is real and positive and $\gamma$ is a positive empirically determined dimensionless coefficient. Thus, the EDQNM for homogeneous turbulence has the considerable simplification and computational efficiency of having an analytical expression for the triad relaxation time $\Theta^{EDQNM}(\mathbf{k},\mathbf{p},\mathbf{q})(t)$, given in Equation (47), with the superscript $0 \to EDQNM$.

For homogeneous turbulence on an $f$–plane, without Rossby waves, including for HIT, $D_\mathbf{k}^0 = \hat{v}_0(k)k^2 > 0$ is real as is $D_\mathbf{k}^\eta(t) \to \mu_\mathbf{k}^{eddy}(t) > 0$ in Equation (47). Thus,

$$\Theta^{EDQNM}(\mathbf{k},\mathbf{p},\mathbf{q})(t) = \frac{1-\exp\left(-\left[\mu_\mathbf{k}(t)+\mu_\mathbf{p}(t)+\mu_\mathbf{q}(t)\right](t-t_0)\right)}{\mu_\mathbf{k}(t)+\mu_\mathbf{p}(t)+\mu_\mathbf{q}(t)} \quad (49)$$

where

$$\mu_\mathbf{k}(t) = \hat{v}_0(k)k^2 + \mu_\mathbf{k}^{eddy}(t) = \hat{v}_0(k)k^2 + \gamma\left[k^2 C_\mathbf{k}(t,t)\right]^{\frac{1}{2}}. \quad (50)$$

Now, $\Theta^{EDQNM}(\mathbf{k},\mathbf{p},\mathbf{q})(t) = \mathrm{Re}\,\Theta^{EDQNM}(\mathbf{k},\mathbf{p},\mathbf{q})(t) \geq 0$ is both real and non-negative and this ensures that the cumulant $C_\mathbf{k}(t,t)$ is also real and non-negative, and thus realizable, as also shown in Appendix B with superscript $EDMAC \to EDQNM$.

For homogeneous anisotropic turbulence interacting with transient Rossby waves on a $\beta$–plane $D_\mathbf{k}^0 = \hat{v}_0(k)k^2 + \mathrm{i}\omega_\mathbf{k}$ and taking $D_\mathbf{k}^\eta(t) \to \mu_\mathbf{k}^{eddy}(t)$ again we have

$$R_\mathbf{k}(t,t') \doteq R_\mathbf{k}^{EDQNM}(t,t') = \exp\left(-[\mu_\mathbf{k}(t)+\mathrm{i}\omega_\mathbf{k}](t-t')\right), \quad (51)$$

and

$$\Theta^{EDQNM}(\mathbf{k},\mathbf{p},\mathbf{q})(t)$$
$$= \frac{1-\exp\left(-\left[\mu_\mathbf{k}(t)+\mu_\mathbf{p}(t)+\mu_\mathbf{q}(t)+\mathrm{i}(\omega_\mathbf{k}+\omega_\mathbf{p}+\omega_\mathbf{q})\right](t-t_0)\right)}{\mu_\mathbf{k}(t)+\mu_\mathbf{p}(t)+\mu_\mathbf{q}(t)+\mathrm{i}(\omega_\mathbf{k}+\omega_\mathbf{p}+\omega_\mathbf{q})}. \quad (52)$$

Unfortunately, the wave terms mean it is not possible to guarantee that $\mathrm{Re}\,\Theta^{EDQNM}(\mathbf{k},\mathbf{p},\mathbf{q})(t) \geq 0$, as noted by Bowman et al. [67], and so there may be situations where $C_\mathbf{k}(t,t)$ is no longer realizable.

*7.2 Analytical Relaxation Functions for EDMAC*

The EDMAC generalizes the EDQNM model by using an analytical form for the triad relaxation function that is realizable in the presence of transient Rossby waves (or drift-waves or indeed other waves). It thus solves the problem that was the focus of Bowman et al. [67] but without the introduction of an auxiliary Markovian evolution equation for the triad relaxation function. As we have noted above, although the auxiliary equation makes the whole system Markovian it is still a large computational overhead since the triad relaxation function depends on time and six spectral space dimensions that reduce to four when the Kronecker delta functions in Equations (35) and (38) are implemented. This is for 2D turbulence with corresponding larger computational effort required for 3D turbulence.

In the EDMAC model for homogeneous anisotropic turbulence interacting with transient Rossby waves on a $\beta$–plane we make the replacement

$$D_\mathbf{k}^r(t) \to \rho_\mathbf{k}(t) + \mathrm{i}\omega_\mathbf{k} \quad (53)$$

where

$$\rho_\mathbf{k}(t) = \mu_\mathbf{k}(t) + c\frac{\omega_\mathbf{k}^2}{\mu_\mathbf{k}(t)} = \hat{v}_0(k)k^2 + \gamma\left[k^2 C_\mathbf{k}(t,t)\right]^{\frac{1}{2}} + c\frac{\omega_\mathbf{k}^2}{\mu_\mathbf{k}(t)} \quad (54)$$

with a typical value of $c = \frac{1}{2}$. Thus, the response function equation becomes

$$R_{\mathbf{k}}(t,t') \doteq R_{\mathbf{k}}^{EDMAC}(t,t') = \exp\left(-[\rho_{\mathbf{k}}(t) + i\omega_{\mathbf{k}}](t-t')\right), \tag{55}$$

and

$$\Theta^{EDMAC}(\mathbf{k},\mathbf{p},\mathbf{q})(t)$$
$$= \frac{1 - \exp\left(-\left[\rho_{\mathbf{k}}(t) + \rho_{\mathbf{p}}(t) + \rho_{\mathbf{q}}(t) + i(\omega_{\mathbf{k}} + \omega_{\mathbf{p}} + \omega_{\mathbf{q}})\right](t-t_0)\right)}{\rho_{\mathbf{k}}(t) + \rho_{\mathbf{p}}(t) + \rho_{\mathbf{q}}(t) + i(\omega_{\mathbf{k}} + \omega_{\mathbf{p}} + \omega_{\mathbf{q}})}. \tag{56}$$

As shown in Appendices A and B, the EDMAC model is realizable for all $c \geq \frac{1}{4}$. The eddy damping in Equation (50) was of course chosen by Orszag [2] on empirical grounds and from a practical point of view it would probably not matter if the frequency renormalized eddy damping was arrived at in the same way. However, in a sequel where we study the performance of the EDMAC model with transient Rossby waves we also aim to arrive at the form above through renormalized perturbation theory. It is clear that for small $\omega_{\mathbf{k}}$ the frequency squared term in Equation (54) is negligible. As well, for large wave numbers $k$, such as in typical geophysical enstrophy cascading inertial ranges, the frequency squared term also becomes negligible. Thus, we expect that the EDMAC model will be as computationally efficient as the EDQNM but will in addition be realizable in the presence of transient Rossby waves (and other waves) since $\operatorname{Re} \Theta^{EDMAC}(\mathbf{k},\mathbf{p},\mathbf{q})(t) \geq 0$ for all $c \geq \frac{1}{4}$. As well, we expect that at least for small amplitude waves the efficient performance and veracity of the EDMAC will be like that of the EDQNM but without the possibility of blow up.

### 7.3 EDMAC for Three-Dimensional Turbulent Flows

There are various possible extensions of the frequency renormalized eddy damping parameterization for the realizable EDMAC that we have presented for interaction of 2D turbulence with Rossby waves. Of course, the EDQNM was initially formulated for 3D HIT [2] and has been applied to a variety of problems in 2D and 3D HIT and HAT including with rotation [46,47]. We expect similar generalizations should be possible for the realizable EDMAC. Indeed, our results are easily extended to the model of 3D quasigeostrophic flows in Appendix B of Frederiksen [60], suitably generalized for flow on a $\beta$-plane with Rossby waves.

## 8. Reflections and Perspectives

### 8.1 Jorgen's Personal Reflections

Jack Herring first loomed large as a giant intellect and pioneer of statistical fluid dynamics during my nine month's visit to NCAR in 1980 as part of a year's sabbatical from CSIRO Division of Atmospheric Physics. I was an early career research scientist at CSIRO having switched from quantum field theory to classical geophysical fluid dynamics five years earlier. My PhD and post doc had involved determining the analytical properties of Feynman diagrams, particularly the class of loop diagrams, which form the basis of dispersion relations. These dispersion relations included the quadratic statistical closure equations that describe meson-meson scattering – the core and iconic problem of strong interaction quantum field theories. I was therefore very keen to learn as much as possible about statistical turbulence closure theory and geophysical fluid dynamics as possible. I suggested that the 'Downunder' tradition of afternoon tea and coffee be instituted at the Climate Dynamics Group and at it, and at lunch, learnt an amazing amount of science from the great scientists at NCAR (a list too long to produce) and from more junior staff and post docs. During this time my friendship with Jack developed and continued over the years.

As documented in Section 2, Jack has been an enormous influence on my career with much of my work following in his footsteps. When the time came to launch into the complexities of the numerical formulation of turbulence closure codes, Jack very generously showed and explained all the tricks needed to efficiently implement the non-Markovian integro-differential equations. Jack was always very friendly, cheerful, and unassuming, always interested, and with perceptive questions including at seminars. Always there during my short and longer visits to NCAR. Jack took a lot of interest, particularly in our work on the inhomogeneous QDIA closures, from being thesis examiner of Terry's PhD on the numerical implementation, and further development of the closure, to our most recent publication in Fluids 2022 [44].

*8.2 Perspectives on Strong Interaction Theories*

The non-Markovian closure theories of Kraichnan, Herring, Edwards and McComb for HIT tackle the core and iconic problem of strong turbulence without the complexities of inhomogeneity. They are accurate at the large energy containing scales but their power law behaviour in inertial ranges have in principle or in practice deficiencies since they do not fundamentally address the vertex renormalization problem. However, restricting the wavenumber ranges in the two-time cumulant and response function equations in one parameter regularized versions of some of these closures gives excellent results. This is also the case for the QDIA closure for inhomogeneous turbulence.

To put into perspective the state of strong interaction turbulence closure theories it is perhaps useful to look at the corresponding situation in strong interaction quantum field theory. As noted in the previous subsection, pion-pion scattering is the core and iconic problem of strong interaction hadron physics. Mandelstam (1958) [87] developed closure equations for hadron scattering which are quadratic in the partial wave spectral scattering amplitudes. A question of major interest was whether the scattering equations have unique solutions that can be calculated through convergent iteration. To answer this question, it was necessary to reformulate the equations, and the proof was established for pion-pion scattering (Frederiksen 1975 [88]; Atkinson et al. 1976 [89]). This was done using functional analysis contraction mapping theorems in Banach space of doubly Hölder continuous scattering amplitudes. However, it has taken until 2023 for the iteration procedure to be numerically implemented to generate the nonperturbative scattering amplitudes in a remarkable effort by Tourkine and Zhiboedov (2023) [90].

While scattering between asymptotic 'in and 'out' states has been the focus of much of strong interaction hadron physics, often with introduction of some empirical data, time-dependent non-equilibrium quantum field theories have become of increasing interest in the last two decades. This is particularly so in studies of Bose-Einstein condensation far from equilibrium [91,92], in studies of cosmology and inflation [93–95], and quark-gluon plasma [96]. In these problems the generalization from homogeneous to inhomogeneous statistical quantum field theory is required to describe the time evolution of quantum fluctuations in spatially inhomogeneous dynamical fields. The Schwinger-Keldysh CTP approach has generally been the basis for these studies. The Schwinger-Keldysh [81,82] equations, like Kraichnan's [97] inhomogeneous DIA (IDIA), and the MSR [53] and path integral extensions [78,79], require the full covariance matrix to close the mean field equation. This is a severe restriction on the size of the problems that can be solved since if the field has $N$ degrees of freedom the full covariance matrix has $N^2$ (the QDIA gets around this problem as described in the Introduction). Consequently, the full inhomogeneous problem has generally only been tackled in one space dimension [98–100]. Interestingly in studies of the evolution of a Bose gas, Cooper et al. [99] found that the bare vertex approximation performs better than other closure schemes like the two-particle irreducible expansion. The bare vertex approximation is of course the basis of the DIA, SCFT, LET and Edward's closures and guarantees the realizability of the DIA, the IDIA, the QDIA and the bare vertex quantum closure equations.

The statistical dynamics of strongly interacting fields is an enormously difficult problem. The achievements of the pioneers in fluid turbulence closure theory have been truly outstanding by any comparison.

## 9. Conclusions

Jack Herring made remarkable pioneering steps and contributions to laying the foundations of the statistical dynamical closure theory of fluid turbulence. Some of his important works have been briefly reviewed, their impacts and related developments discussed, and some of the further developments and extensions summarized.

A particular focus of this article has also been to present the eddy Damped Markovian Anisotropic Closure (EDMAC), a generalization of the Eddy Damped Quasi Normal Markovian (EDQNM) that is realizable for anisotropic turbulence interacting with transient waves such as Rossby waves. By construction the EDMAC is as computationally efficient as the EDQNM but overcomes a long-standing problem of realizability with transient waves present. This builds on Herring's (1975) [1] generalization of the EDQNM for anisotropic 2D turbulence and his studies of the relaxation to isotropy.

**Author Contributions:** Conceptualization, JSF and TJO; methodology, JSF; formal analysis, JSF; investigation, JSF and TJO; writing—original draft preparation, JSF; writing—review and editing, JSF and TJO; supervision, JSF; project administration, JSF; funding acquisition, TJO. All authors have read and agreed to the published version of the manuscript.

**Funding:** TJO was funded by CSIRO Environment.

**Data Availability Statement:** Data sharing is not applicable since no new data were created or analysed in this article.

**Conflicts of Interest:** The authors declare no conflict of interest.

## Appendix A: Positive Semi-definite Triad Relaxation Time with Rossby Waves

As noted in Section 7, lack of realizability of the EDQNM model in the presence of transient waves can be overcome by using a frequency renormalized eddy damping. The response function is then:

$$R_{\mathbf{k}}(t,t') = \exp-[\rho(t)_{\mathbf{k}} + i\omega_{\mathbf{k}}](t-t') \tag{A1}$$

where $\rho_{\mathbf{k}}(t) = \mu_{\mathbf{k}}(t) + c\omega_{\mathbf{k}}^2/\mu_{\mathbf{k}}(t) > 0$ is the frequency renormalized damping, $\mu_{\mathbf{k}}(t) > 0$ is the damping, $\omega_{\mathbf{k}}$ is the Rossby wave frequency and a typical value of $c = \frac{1}{2}$. The triad relaxation time is then given by

$$\Theta^{EDMAC}(\mathbf{k}_1, \mathbf{k}_2, \mathbf{k}_3)(t) = \frac{1 - \exp-[\rho + i\omega](t-t_0)}{\rho + i\omega} \tag{A2}$$

where, without loss of generality, we take $t_0 = 0$. Here,

$$\begin{aligned} \mu &= \mu_{\mathbf{k}_1}(t) + \mu_{\mathbf{k}_2}(t) + \mu_{\mathbf{k}_3}(t), \\ \omega &= \omega_{\mathbf{k}_1} + \omega_{\mathbf{k}_2} + \omega_{\mathbf{k}_3} \\ \rho_j &= \mu_{\mathbf{k}_j} + c\frac{\omega_{\mathbf{k}_j}^2}{\mu_{\mathbf{k}_j}}, \quad j = 1, 2, 3 \\ \rho &= \rho_1 + \rho_2 + \rho_3. \end{aligned} \tag{A3}$$

Then

$$\operatorname{Re}\Theta^{EDMAC}(\mathbf{k}_1,\mathbf{k}_2,\mathbf{k}_3)(t)$$
$$= \frac{1}{\rho^2+\omega^2}\left[\rho\{1-[\exp-\rho t]\cos\omega t\}+\omega[\exp-\rho t]\sin\omega t\right] \quad (A4)$$

where

$$\frac{\partial \operatorname{Re}\Theta^{EDMAC}(\mathbf{k}_1,\mathbf{k}_2,\mathbf{k}_3)(t)}{\partial t} = [\exp-\rho t]\cos\omega t. \quad (A5)$$

Next, we examine the conditions under which $\operatorname{Re}\Theta^{EDMAC}(\mathbf{k}_1,\mathbf{k}_2,\mathbf{k}_3)(t)\geq 0$ so that the EDMAC model of Section 7 (Equations (39) and (43)) is realizable as outlined in Appendix B. Firstly, we note that for $t\geq 0$, $\cos\omega t$ is then an even function of $\omega$ and both $\omega$ and $\sin\omega t$ are odd functions with their product $\omega\sin\omega t$ being even. Thus, $0<|\omega|t\leq\pi$

$$1-[\exp-\rho t]\cos|\omega|t>0 \ ; \ |\omega|\,[\exp-\rho t]\sin|\omega|t\geq 0. \quad (A6)$$

Further, for $|\omega|t>\pi$,

$$\exp-\rho t < \exp-(\rho\frac{\pi}{|\omega|}) < \exp-(\pi) < 0.05 \quad (A7)$$

provided

$$\rho \geq |\omega|. \quad (A8)$$

Of course, in general

$$-1\leq\cos|\omega|t\leq 1;\ -1\leq\sin|\omega|t\leq 1 \quad (A9)$$

and thus for $|\omega|t>\pi$,

$$\left[\rho^2+\omega^2\right]\operatorname{Re}\Theta^{EDMAC}(\mathbf{k}_1,\mathbf{k}_2,\mathbf{k}_3)(t) > 0.9\rho > 0. \quad (A10)$$

The relationship in Equation (A8) holds if

$$\left(\mu_{\mathbf{k}_j}(t)+c\frac{\omega_{\mathbf{k}_j}^2}{\mu_{\mathbf{k}_j}(t)}\right)\geq|\omega_{\mathbf{k}_j}|,\ j=1,2,3 \quad (A11)$$

and the inequalities in Equation (A11) are valid provided $c\geq \frac{1}{4}$ which is established by solving simple quadratic equations.

**Appendix B: Langevin Equation for EDMAC model**

The EDMAC model in Equations (39) and (43) can be shown to be realizable since it is underpinned by a stochastic model as is the EDQNM as discussed by Leith [3] and Herring and Kraichnan [66]. The Langevin equation which allows precise construction of the EDMAC model is given by:

$$\left(\frac{\partial}{\partial t}+D_{\mathbf{k}}^0+D_{\mathbf{k}}^\eta(t)\right)\tilde{\zeta}_{\mathbf{k}}(t)=f_{\mathbf{k}}^{\ 0}(t)+f_{\mathbf{k}}^{\ S}(t) \quad (A12)$$

where $D_{\mathbf{k}}^0$ is given in Equation (34) and $D_{\mathbf{k}}^\eta(t)$ is given in Equation (53) with $X=0$. As well,

$$f_{\mathbf{k}}^{\ 0}(t)=\tilde{f}_{\mathbf{k}}^o, \quad (A13)$$

and

$$f_{\mathbf{k}}^{\ S}(t)=\sqrt{2}\sum_{\mathbf{p}}\sum_{\mathbf{q}}\delta(\mathbf{k}+\mathbf{p}+\mathbf{q})K(\mathbf{k},\mathbf{p},\mathbf{q})\left[\operatorname{Re}\Theta^{EDMAC}(\mathbf{k},\mathbf{p},\mathbf{q})(t)\right]^{\frac{1}{2}}$$
$$\times w(t)\rho_{-\mathbf{p}}^{(1)}(t)\rho_{-\mathbf{q}}^{(2)}(t). \quad (A14)$$

The variables $\rho_{\mathbf{k}}^{(i)}(t)$, where $i$=1, 2 or 3, and $w(t)$ are independent random variables that satisfy the following relationships:

$$< \rho_{\mathbf{k}}^{(i)}(t)\rho_{-\mathbf{l}}^{(j)}(t') >= \delta_{ij}\delta_{\mathbf{kl}}C_{\mathbf{k}}(t,t'), \tag{A15}$$

with

$$< \tilde{\zeta}_{\mathbf{k}}(t)\tilde{\zeta}_{-\mathbf{k}}(t') >= C_{\mathbf{k}}(t,t'), \tag{A16}$$

and

$$< w(t)w(t') >= \delta(t-t'). \tag{A17}$$

Here, $\delta$ is the Kronecker delta function in Equation (A15), and in Equation (A17) it is the Dirac delta function.

The realizability of the cumulants $C_{\mathbf{k}}(t,t)$, in the EDMAC model is established by the Langevin equation provided $\text{Re}\,\Theta^{EDMAC}(\mathbf{k},\mathbf{p},\mathbf{q})(t) \geq 0$. This is in turn is shown in Appendix A to be the case provided $c \geq 1/4$. The EDMAC equations also preserve conservation of kinetic energy and potential enstrophy (in the absence of forcing and dissipation).